\documentclass[twocolumn,floats,floatfix,amssymb,nofootinbib,prd,superscriptaddress,aps]{revtex4-2}
\usepackage{graphicx,amssymb,amsmath,amsthm,amsfonts,epsfig}
\usepackage[linktocpage,colorlinks=true,citecolor=OliveGreen,linkcolor=Maroon,urlcolor=Maroon]{hyperref}
\usepackage[usenames]{color}
\usepackage{epstopdf}
\usepackage{bm}
\usepackage{dcolumn}
\usepackage[utf8]{inputenc}
\usepackage{latexsym}
\usepackage{rotating}
\usepackage{hyperref}
\usepackage{tabularx}
\usepackage{braket}
\usepackage{color}
\usepackage{esvect}
\usepackage{longtable}
\usepackage{enumerate}
\usepackage{footmisc}
\usepackage{array}
\usepackage{tensor}
\usepackage{mathtools}
\usepackage{url}
\usepackage[dvipsnames]{xcolor}
\usepackage{comment}
\usepackage{multirow}
\usepackage{nicematrix}
\renewcommand{\vec}[1]{\boldsymbol{#1}}

\newcommand{\ben}{\begin{enumerate}}
\newcommand{\een}{\end{enumerate}}

\def\be{\begin{equation}}
\def\ee{\end{equation}}
\def\bea{\begin{eqnarray}}
\def\eea{\end{eqnarray}}
\def\beq{\begin{eqnarray}}
\def\eeq{\end{eqnarray}}
\newcommand{\ba}{\begin{align}}
\newcommand{\ea}{\end{align}}
\def\ba{\bar{a}}

\definecolor{cornellGreen}{HTML}{6EB43F}
\definecolor{cornellRed}{HTML}{B31B1B}

\begin{document}
\title{Black Hole Spectroscopy in Environments:~Detectability Prospects}
\author{Thomas F.M.~Spieksma}
\affiliation{Center of Gravity, Niels Bohr Institute, Blegdamsvej 17, 2100 Copenhagen, Denmark}
\author{Vitor Cardoso}
\affiliation{Center of Gravity, Niels Bohr Institute, Blegdamsvej 17, 2100 Copenhagen, Denmark}
\affiliation{CENTRA, Departamento de F\'{\i}sica, Instituto Superior T\'ecnico -- IST, Universidade de Lisboa -- UL,
Avenida Rovisco Pais 1, 1049 Lisboa, Portugal}
\author{Gregorio Carullo}
\affiliation{Center of Gravity, Niels Bohr Institute, Blegdamsvej 17, 2100 Copenhagen, Denmark}
\affiliation{School of Physics and Astronomy and Institute for Gravitational Wave Astronomy, University of Birmingham, Edgbaston, Birmingham, B15 2TT, United Kingdom}
\author{Matteo Della Rocca}
\affiliation{Dipartimento di Fisica, Universit\`a di Pisa, Largo B. Pontecorvo 3, 56127 Pisa, Italy}
\affiliation{INFN, Sezione di Pisa, Largo B. Pontecorvo 3, 56127 Pisa, Italy}
\author{Francisco Duque}
\affiliation{Max Planck Institute for Gravitational Physics (Albert Einstein Institute) 
Am Mühlenberg 1, D-14476 Potsdam, Germany}
\begin{abstract}
The ringdown phase following a binary black hole coalescence is a powerful tool for measuring properties of the remnant black hole. Future gravitational wave detectors will increase the precision of these measurements and may be sensitive to the environment surrounding the black hole. This work examines how environments affect the ringdown from a binary coalescence. Our analysis shows that for astrophysical parameters and sensitivity of planned detectors, the ringdown signal is indistinguishable from its vacuum counterpart, suggesting that ringdown-only analyses can reliably extract the (redshifted) mass and spin of the remnant black hole. These conclusions include models with spectral instabilities, suggesting that these are not relevant from an observational viewpoint. Deviations from inspiral-only estimates could then enhance the characterisation of environmental effects present during the coalescence.
\end{abstract}
\maketitle
%

\noindent {\bf \em Introduction.} 
Albeit modeling of gravitational-wave (GW) signals from black hole (BH) mergers is most often carried out in vacuum, the Universe is permeated with various forms of matter, such as dark matter or interstellar dust.
In the vicinity of BHs, matter can accumulate, leading to the formation of ``environments'' like dark matter halos or accretion disks. The impact of these structures on the GW emission from BH binaries, and the potential to reveal properties of the constituent matter through GWs is an active area of research~\cite{Kocsis:2011dr,Barausse:2014tra,Barack:2018yly, Annulli:2020ilw,Garg:2022nko,Cardoso:2022whc, Derdzinski:2020wlw, Zwick:2021dlg, Cardoso:2021vjq,Speri:2022upm,Gupta:2022fbe,DeLuca:2022xlz,Cole:2022yzw,Rahman:2023sof,Kejriwal:2023djc,Tomaselli:2023ysb, Brito:2023pyl,Duque:2023seg, Tomaselli:2024bdd,Tomaselli:2024dbw,Cannizzaro:2024fpz}.

In the last stages of a binary BH coalescence, the progenitors merge and form a dynamical remnant BH, which then relaxes to a quasi-stationary state~\cite{Buonanno:2006ui,Berti:2007fi}.
Past an initial transient phase, the GW signal during relaxation is well-described by a superposition of exponentially damped sinusoids whose frequencies correspond to the characteristic ``quasi-normal modes'' (QNMs) of the final BH~\cite{Kokkotas:1999bd,Berti:2009kk,Konoplya:2011qq}. The simplicity of the signal provides insights into the properties of the remnant, sparking the development of the ``BH spectroscopy program'' (see, e.g.,~\cite{Baibhav:2023clw} and references therein). 
In vacuum, the identification of a single QNM frequency allows to determine the mass and spin of the remnant BH, while measurement of more than one QNM enables consistency tests of General Relativity~\cite{Berti:2005ys,Berti:2016lat,Baibhav:2023clw}. But what exactly is the impact of astrophysical environments on BH spectroscopy? Do they affect the \emph{detectability} of a signal? Can we \emph{distinguish} them from the pure-vacuum case?

Astrophysical toy models with charged BHs have recently shown that the fundamental QNM frequency can shift significantly~\cite{Cannizzaro:2024yee}, a departure observed even in time-domain evolutions. A comprehensive set of works have established that QNMs are in general ``spectrally unstable'' against small perturbations in the underlying spacetime~\cite{Nollert:1996rf,Nollert:1998ys,Leung:1997was,Barausse:2014tra,Jaramillo:2021tmt,Cardoso:2019rvt,Jaramillo:2020tuu,Cheung:2022rbm,Konoplya:2022pbc,Konoplya:2022hll,Cardoso:2022whc,Cardoso:2024mrw,Yang:2024vor,Ianniccari:2024ysv}, which might correspond to the one caused by astrophysical environments. However, 
time-domain analyses in these same geometries showed that the prompt, early-time ringdown signal is not affected by spectral instabilities, questioning its relevance for GW astronomy~\cite{Berti:2022xfj,Cardoso:2024mrw,Yang:2024vor}. Nonetheless, from a data analysis point of view, no quantification of the effect of realistic environments in BH spectroscopy has ever been attempted.\footnote{This statement concerns the dominant mode and especially higher overtones, which are generically afflicted by stronger instabilities~\cite{Cardoso:2024mrw}. While data-analysis oriented studies exist~\cite{Leong:2023nuk,Redondo-Yuste:2023ipg,DeLuca:2024uju}, they focus on ad hoc matter profiles, or accreting Vaidya spacetimes, less relevant from an astrophysical viewpoint.}

In this work, we investigate whether the ringdown signal in the presence of a realistic BH environment can be distinguished from its vacuum counterpart. We focus on highly asymmetric binaries, for which the environment is expected to survive the inspiral phase. This is a conservative assumption, since environments of comparable mass binaries are significantly more depleted~\cite{Aurrekoetxea:2024cqd}. We consider the heavier binary component as a BH at the center of a galaxy, surrounded by a halo (of dark or baryonic matter). This class of sources is a prime target for future space-based GW detectors, such as LISA~\cite{LISA:2017pwj,Baker:2019nia}. Our procedure intends to be agnostic regarding the nature of a possible instability, using only the well-understood vacuum waveform as a ruler to measure how well we can differentiate environmental effects in the ringdown.

\noindent {\bf \em Dirty black holes.} 
As a proxy for the galactic environment, we consider a BH at the center of some ``halo'' matter distribution. For simplicity, we take the spacetime to be spherically symmetric with line element 
\begin{equation}\label{eqn:line_element}
ds^2= -A(r)\mathrm{d}t^2 + \frac{\mathrm{d}r^2}{1-2m(r)/r} + r^2 \mathrm{d}\Omega^2\,,
\end{equation}
where $\mathrm{d}\Omega^2 = \mathrm{d}\theta ^2 + \sin \theta^2 \mathrm{d}\varphi^2 $ is the line element of the $2$--sphere and $m(r)$ is the mass function of the system. A systematic approach to construct stationary solutions for these systems is the ``Einstein cluster'' construction, which takes a collection of particles in all possible circular geodesics~\cite{Einstein_cluster,Einstein_cluster_2}. The effective energy-momentum tensor is equivalent to  an anisotropic fluid with vanishing radial pressure, and only tangential pressure $p_{\rm t}$,
\begin{equation}
T^{\mu}_{\nu} = \mathrm{diag}(-\rho, 0, p_{\rm t}, p_{\rm t})\,.
\end{equation}
This method was used to study binaries in the presence of an environment with the particular mass function~\cite{Cardoso:2021wlq, Cardoso:2022whc} 
\begin{equation}
m(r)=M_{\mathrm{BH}}+\frac{M_{\rm H} r^2}{\left(a_{\rm H}+r\right)^2}\left(1-\frac{2 M_{\mathrm{BH}}}{r}\right)^2\,, \label{eq:HernquistMass}
\end{equation}
where $M_{\rm H}, a_{\rm H}$ are, respectively, the mass and characteristic length of the halo and $A(r)$ in the metric~\eqref{eqn:line_element} is given explicitly in~\cite{Cardoso:2021wlq}. This profile is dominated by the BH gravity at $r \ll a_{\rm H}$ and at large distances recovers the mass profile of the Hernquist model, which describes elliptical galaxies and galactic bulges~\cite{1990ApJ...356..359H, Quinlan:1994ed}. 
We also define the compactness and density of the halo as
\begin{equation}\label{eq:compactness}
\mathcal{C} = \frac{M_{\rm H}}{a_{\rm H}}\quad \text{and} \quad \rho\sim \frac{{\cal C}^3}{M_{\rm H}^2}\,,
\end{equation}
respectively. Both quantities affect the GW-response of the system. Halos are expected to be much more massive than the central BH they host, i.e., $M_{\rm BH} \ll M_{\rm H}$, and have low compactness ($\mathcal{C}\lesssim 10^{-4}$). Other astrophysical environments, such as boson clouds composed by ultralight fields, can have much higher compactness (and density)~\cite{Brito:2015oca}. The Hernquist profile is one of many possible choices, and it is straightforward to repeat the same procedure to find generic stationary spacetimes describing BHs dressed by matter~\cite{Figueiredo:2023gas,Speeney:2024mas}. Yet, all of these exhibit the same qualitative behavior and therefore we focus on the mass function in Eq.~\eqref{eq:HernquistMass}, treat both $M_{\rm H}$ and $a_{\rm H}$ as free parameters, and take it as a proxy for generic distributions of matter around a BH. 

We consider a barotropic equation of state, for which changes in pressure, $\delta p_{\rm t, r}$ (tangential and radial), and density $\delta \rho$, are related by the speed of sound:
\begin{equation}\label{eqn:speedofsound1}
\delta p_{\rm t, r}=c_{\rm s_{t, r}}^2 \delta \rho\,, \quad \text{with} \quad c_{\rm s,r} = \left(\frac{2M_{\rm BH}+a_{\rm H} }{r+a_{\rm H}}\right)^{4}\,.
\end{equation}
Small sound speeds lead to problems regarding the well posedness of the system~\cite{Schoepe:2017cvt}, making it challenging to solve numerically. Following~\cite{Cardoso:2022whc,Datta:2023zmd,Speeney:2024mas}, we choose the \textit{ad hoc} profile in Eq.~\eqref{eqn:speedofsound1}, and from previous works we do not expect major qualitative changes for other profiles.

\noindent {\bf \em Methods.} 
We perturb the spacetime~\eqref{eqn:line_element} by plunging a particle into the BH~\cite{Regge:1957td,Zerilli:1970wzz,Zerilli:1970se, Barack:2018yvs, Pound:2021qin, Cardoso:2022whc, Duque:2023seg,Cardoso:2021wlq, Zenginoglu:2011zz}. Without loss of generality, the plunge is along the radial $\hat{z}$--direction, exciting only axially symmetric polar modes. The corresponding waveform is computed using recently developed techniques of BH perturbation theory in nonvacuum (spherically symmetric) spacetimes~\cite{Barack:2018yvs, Pound:2021qin,Cardoso:2021wlq,Cardoso:2022whc,Duque:2023seg}. Further details are found in Appendix~\ref{sec:numeric_procedure}. After the plunge, the GW signal is well-described by
a superposition of exponentially damped sinusoids:\footnote{As the polarization axes are oriented along the $\theta$ and $\varphi$--direction, the cross-polarization ($h_{\times}$) is zero and the GW radiation is purely captured by $h_+$.}
\begin{equation}
h_+\!=\!\frac{M_{\rm BH}}{r}\,\mathrm{Re}\!\left[\,\sum_{n = 0}^{\infty}\sum_{\ell = 2}^{\infty} A_{\ell 0 n}\!\times\! e^{-i (\omega_{\ell 0 n} t-\phi_{\ell 0 n})}\!{}_{{\scalebox{0.65}{$-$}}2}\mkern-2mu Y_{\ell 0}(\theta, \varphi)\right]\,,
\label{eq:hplus_ringdown}
\end{equation}
where $h_+=h_+(t,r,\theta,\varphi)$, ${}_{{\scalebox{0.65}{$-$}}2}\mkern-2mu Y_{\ell 0}(\theta,\varphi)$ are spin-weighted spherical harmonics and $\omega_{\ell 0 n} = \omega_{\mathrm{R},\ell 0 n}+i\omega_{\mathrm{I},\ell 0 n}$ are the QNM frequencies. For radial plunges and our choice of axis, $m=0$ in Eq.~\eqref{eq:hplus_ringdown}. This signal corresponds to light-ring relaxation~\cite{Cardoso:2019rvt}. In vacuum, it eventually gives way, at late times, to power-law tails from curvature backscattering, either when considering vacuum perturbations~\cite{Price:1971fb,Leaver:1986gd,Gundlach:1993tp,Hintz:2020roc} or infalling particles~\cite{DeAmicis:2024not,Albanesi:2023bgi,Carullo:2023tff,Cardoso:2024jme,Islam:2024vro}, but the structure in the presence of surrounding matter is richer, as we will see below.

We expect astrophysical setups with a hierarchy of scales $M_{\rm BH} \ll M_{\rm H} \ll a_{\rm H}$. These scales and the need to extract the signal far away from the system pose a challenge and restrict the range of halo mass and size that we can study. We thus focus on relatively large halo compactnesses. As we will show, this choice leads to an {\it overestimate} of the impact of the environment, which strengthens our conclusions. Specifically, we take $M_{\rm H} = \left[0.1,0.3,0.5,1,10\right]M_{\rm BH}$, while varying $a_{\rm H}$, ensuring the GW signal is always extracted \emph{outside} the halo, $r_{\rm ex} \gg a_{\rm H}$. To avoid errors associated with finite extraction radii, we extract the wave at different finite radii and extrapolate the signal to infinity, by fitting a polynomial $r h(t) = h_{\infty} + a_{1}/r + a_{2}/r^2 + \cdots$, where $h_{\infty}$ is the waveform at $\mathcal{I}^{+}$. In all our simulations, the particle starts the plunge at $r_{\rm p}(t=0) = 100M_{\rm BH}$, and we extract the signal at $r_{\rm ex} = \left[20,40,80,100\right]a_{\rm H}$. 
As the trajectory of the particle depends on the ``halo density'' it encounters, i.e., $\mathrm{d}t/\mathrm{d}\tau \propto 1/\sqrt{A(r)}$ (see Eq.~\eqref{eqn:line_element}, where $\tau$ is the proper time), the strain amplitude and instant at which the waveform peaks do depend on the choice of compactness. 
Since searches will only have access to the redshifted BH mass, we allow waveforms to be stretched (accounting for the gravitational redshift) and shifted when comparing signals from different configurations, in order to maximize the faithfulness (cf.~Eq.~\eqref{eq:faithfulness}), following realistic GW searches. Details are provided in Appendix~\ref{sec:waveform_alignment}. We find significant evidence that GWs are indeed redshifted as they propagate outwards.

To understand if the ringdown of a BH surrounded by an astrophysical environment can be distinguished from its vacuum counterpart, we compute the \emph{faithfulness} between two waveforms $h_{1}$ and $h_{2}$, defined as
\begin{equation}\label{eq:faithfulness}
\mathcal{F} \equiv \mathop{\text{max}}_{t_{\rm c},\phi_{\rm c}} \frac{(h_1|h_2)}{\sqrt{(h_1|h_1)(h_2|h_2)}}\,,
\end{equation}
where $t_{\rm c}$ and $\phi_{\rm c}$ are, respectively, time and phase of the signal at the coalescence. The inner product $(h_1|h_2)$ is
\begin{equation}\label{eq:inner_product}
(h_1|h_2) = 4\,\mathrm{Re} \int_{0}^{\infty} \frac{\tilde{h}_{1}(f)\tilde{h}^{*}_{2}(f)}{S_{n}(f)} \mathrm{d}f\,,
\end{equation}
where tildes indicate a Fourier transform and $S_{n}(f)$ is the one-sided power spectral density, which depends on the specific detector. For reference, we consider the LISA sensitivity curve~\cite{Robson:2018ifk}. 
We maximize over time and phase, with phase maximization done by taking the absolute value instead of the real part in Eq.~\eqref{eq:faithfulness}~\cite{Owen:1995tm}. We always take $h_{2}$ to be the vacuum waveform, which serves as the fiducial signal to compare against. Finally, we define the signal-to-noise ratio (SNR) as 
\begin{equation}
\mathrm{SNR}^{2} = 4\int_{0}^{\infty}\frac{\tilde{h}^{*}(f)\tilde{h}(f)}{S_{n}(f)}\mathrm{d}f\,.
\end{equation}

If two waveforms fulfill the criterion:
\begin{equation}\label{eq:criterion}
    1-\mathcal{F} < \frac{D}{2\,\mathrm{SNR}^{2}}\,,
\end{equation}
for a certain choice of $S_{n}(f)$ and respective SNR, they are classified as \emph{indistinguishable}~\cite{Flanagan:1997kp,Lindblom:2008cm,McWilliams:2010eq,Chatziioannou:2017tdw,Purrer:2019jcp} in the sense that the deviation between two waveforms $\delta h$ satisfies $\braket{\delta h|\delta h} < 1$ (see Eqs.~(8.1)--(8.2)--(8.13) in~\cite{Flanagan:1997kp}). Here, $D$ denotes the dimension of the parameter space one considers, amounting to e.g. $D \sim 10$ for a two damped sinusoids analysis. Note that this criterion formally holds in the limit of high SNR which, with increased sensitivity of future detectors, is a well-justified assumption. 

\noindent {\bf \em The GW signal.}
%
\begin{figure}
    \centering
    \includegraphics[width = 0.48\textwidth]{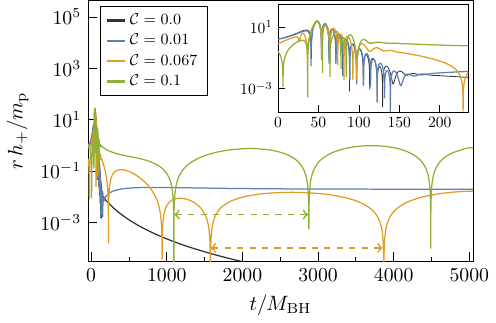}	
    \caption{GW signal $h_+$ when a pointlike particle collides with a BH ``dressed'' by a halo of varying compactness and mass $M_{\rm H} = 10M_{\rm BH}$. Particle begins from rest at $r_{\rm p}= 100M_{\rm BH}$ and the signal is extracted at $r_{\rm ex} = 3000M_{\rm BH}$. Vacuum signal is shown in black. All waveforms are aligned in time and amplitude such that peak strain is at $t = 0$. The oscillations in the orange and green lines indicate the presence of a ``fluid-driven'' mode, imprinted on the GW signal. Its period is well approximated by $T\sim a_{\rm H}/c_{\rm s,r}$ (using Eq.~\eqref{eqn:speedofsound1} with $r\sim a_{\rm H}$). Inset shows zoom-in of the prompt ringdown.}
    \label{fig:Plunge_fluid}
\end{figure}
Consider a binary of mass ratio $q = m_{\rm p}/M_{\rm BH}$, where $m_{\rm p}$ is the mass of the smaller body. We excite the ringdown with a simple head-on collision. Figure~\ref{fig:Plunge_fluid} shows the time evolution of the GW strain when a particle collides with a BH in the presence of a halo with different compactnesses. The main features are:~(i) the early-time, dominant component is a prompt ringdown stage corresponding to light-ring excitation and trapping. This stage is very similar for different halo compactnesses, but {\it is} affected by the environment (for ${\cal C}=0.1$ the signal is noticeably different). By lowering compactness, the signal tends towards the vacuum one (black line);
~(ii) after the prompt ringdown, halo modes set in (on scales set by $a_{\rm H}$) and dominate the late-time signal. This contribution is ``fluid driven'' (seen before in~\cite{Cardoso:2022whc}) and originates from the coupling between the matter and gravity sector. The amplitude and frequency of halo modes depend on the compactness. We find that they oscillate with period $T \sim a_{\rm H}/c_{\rm s,r}$~\cite{Cardoso:2022whc} for all setups studied. For ${\cal C}=0.1, M_{\rm H}=10M_{\rm BH}$ for example, it corresponds to $a_{\rm H}=100M_{\rm BH}$, $T\sim 1500 M_{\rm BH}$, even when evolving the system for longer timescales than shown in Fig.~\ref{fig:Plunge_fluid}. We expect a power-law tail on yet larger timescales, which our current numerical infrastructure cannot probe; 
~(iii) keeping halo compactness fixed while varying halo mass and size, we find that compactness is the dominant factor determining changes in the ringdown signal with respect to vacuum, at least for the parameter space probed.

\begin{figure}
    \centering
    \includegraphics[width = 0.5\textwidth]{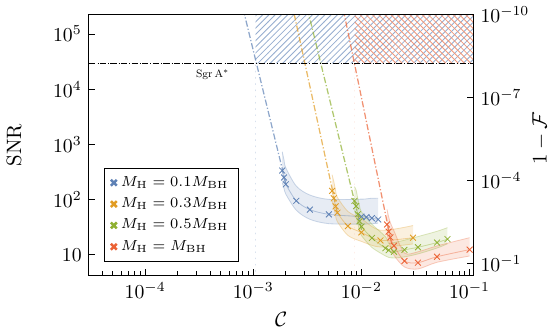}	
    \caption{$\mathrm{SNR}$ required to distinguish the ringdown signal from vacuum for different compactnesses and a given value of halo mass, using the criterion~\eqref{eq:criterion} and $D = 10$. 
    For a given SNR (determining an horizontal line) and halo mass curve, only the region of compactness right of the intersection between the horizontal line and the curve is distinguishable from vacuum.
    The black horizontal line represents a putative signal from $\mathrm{Sgr}\,\mathrm{A}^{*}$ with $q = 10^{-5}$~\eqref{eq:SNR_analytic}. 
    Given its $\mathrm{SNR}$, the signal can be distinguished from vacuum for high compactness:~the blue (red) diagonal-line region for $M_{\rm H} = 0.1M_{\rm BH}\,(M_{\rm BH})$. 
    The right axis shows the corresponding mismatch value (without assuming a value for $D$). 
    Colored dash-dotted lines are a power-law fit through the last few points.
    The error in the shaded regions comes from extrapolating the waveform keeping $1/r$ and up to $1/r^3$ terms.}
    \label{fig:Detectability_fluid}
\end{figure}

\noindent {\bf \em Faithfulness of vacuum templates.}
Figure~\ref{fig:Detectability_fluid} shows the faithfulness~\eqref{eq:faithfulness} for varying halo mass and compactness. 
Faithfulness approaches unity in the limit of zero compactness, even if the total halo mass is large:~the signal is simply redshifted to lower frequencies and a vacuum template will bias the BH mass, which is confirmed in a frequency-domain approach~\cite{Pezzella:2024tkf}. Indeed, for e.g.~$M=M_{\rm BH}, a_{\rm H}=58M_{\rm BH}$, our best match is achieved for a redshift $\alpha=0.978$, compared to the ``expected'' redshift $\sim 1-M_{\rm H}/a_{\rm H}=0.982$.
For small $\mathcal{C}$, the mismatch $1-\mathcal{F}$ decreases as a power-law. While determining the precise scaling analytically would be insightful, waveform stretching leads to some loss of information. Nevertheless, it can be shown that when perturbations of the potential scale as $\mathcal{C}^n$, the faithfulness scales as $\mathcal{C}^{2n}$. 
Counterintuitively, lower halo masses thus allow to probe a larger range of compactness. This can be explained from Eq.~\eqref{eq:compactness}:~for fixed compactness, lower halo masses actually increase the density of the halo close to the BH, making the impact on the ringdown more severe.
However, for low enough halo masses the rise of the plateau (displayed by the curves at high compactness) will prevent the environment discrimination, as intuitively expected.
For large compactnesses, the results showcase a complex behavior: as we allow the template to redshift, the result displays nontrivial features at large densities and compactnesses.

To understand whether the ringdown signal can be distinguished from vacuum, the faithfulness should be compared against the expected $\mathrm{SNR}$ from astrophysical events~\eqref{eq:criterion}. Taking LISA as an example, using Eq.~(3.12b) in~\cite{Berti:2005ys} and Eq.~(4.12) in~\cite{Berti:2007fi}, we find
\begin{equation}\label{eq:SNR_analytic}
\begin{aligned}
    \mathrm{SNR} &= 2.9\times 10^{4} \left(\frac{q}{10^{-5}}\right)\left(\frac{M_{\rm BH}}{4.8\times 10^{6}M_{\odot}}\right)^{3/2}\\& \times\left(\frac{8.3\,\mathrm{kpc}}{D_{\rm L}}\right)
    \left(\frac{3.8
    \times 10^{-40}\,\mathrm{Hz}^{-1}}{S_{n}(f)}\right)^{1/2}\,,
\end{aligned}
\end{equation}
where we consider the fundamental mode $M_{\rm BH}\,\omega_{200} = 0.374$, typical values for the LISA sensitivity curve~\cite{Robson:2018ifk}, and we take $\mathrm{Sgr}A^{*}$ as reference, with mass $M_{\mathrm{Sgr}\,\mathrm{A}^{*}} = 4.8\times 10^{6}M_{\odot}$ and located at $D_{\rm L} = 8.3\,\mathrm{kpc}$~\cite{2008ApJ...689.1044G,2019A&A...625L..10G,GRAVITY:2021xju}.\footnote{Eq.~\eqref{eq:SNR_analytic} agrees with Eq.~(3.21) in~\cite{Berti:2005ys} for equal-mass binaries and their (outdated) LISA noise curve. Note however, that the dependence on mass ratio in~\eqref{eq:SNR_analytic} is $q/(1+q)^2$, which becomes relevant only for equal masses.} Using Eq.~\eqref{eq:criterion} with $D = 10$, we find that when $1-\mathcal{F} < 5.9\times 10^{-9}$ the signal cannot be distinguished from vacuum for the benchmark parameters in~\eqref{eq:SNR_analytic}. From Fig.~\ref{fig:Detectability_fluid}, we see that, for e.g.~$M_{\rm H} = 0.1M_{\rm BH}$, any halo with $\mathcal{C} \lesssim \mathcal{O}(10^{-3})$ is indistinguishable using ringdown, even in the overly optimistic scenario of a signal from $\mathrm{Sgr}A^{*}$. Note that mergers of supermassive BHs are predicted to be exceptionally loud but also more distant, resulting in similar or lower SNR~\cite{Bhagwat:2021kwv}. As part of the surrounding environment could be depleted in such mergers, the scenario we consider is expected to be conservative, suggesting our conclusions hold even in this case.

There are two interesting applications of our results in the context of the Milky Way. The GRAVITY Collaboration constrained the mass within the orbit of the S2 star (highly eccentric, we take it to have radius $\sim 10^4M$) to be $\lesssim 10^{-3} M$, using a Plummer profile or one appropriate for bosonic bound states~\cite{GRAVITY:2021xju,GRAVITY:2023cjt}. From the $M_{\rm H}$ scaling of our results, we find that for a collision at the center of our galaxy to discriminate an environment via ringdown, then $a_{\rm H}\lesssim 10^2 M_{\rm BH}$. In addition, at larger scales, the Milky Way bulge stellar mass is of order $(2.0 \pm 0.3)\times  10^{10}M_\odot$ with a redshift of order $10^{-6}$~\cite{2016A&A...587L...6V}. Figure~\ref{fig:Detectability_fluid} leaves little room for doubt:~BH spectroscopy will not be able to inform us on physics at these scales.

Sources farther away will have a $\mathrm{SNR}$ too low to distinguish any value of the compactness from the vacuum waveform. Since realistic galactic halos have $\mathcal{C}\lesssim 10^{-4}$~\cite{Navarro:1995iw,Kim:2004tc}, we conclude that environments cannot be distinguished with ringdown, using currently planned detectors.

\noindent {\bf \em  Spectral instability.}
%
\begin{figure}
    \centering
    \includegraphics[width = 0.45\textwidth]{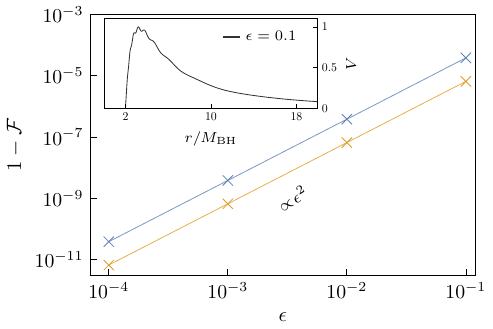}	
    \caption{Mismatch for various values of $\epsilon$, which quantifies the modification to the potential~\eqref{eq:perturbed_potential}. $1-\mathcal{F}$ goes down exactly as $\epsilon^{2}$, showcasing that mode instabilities do not affect observations to any significant degree. Blue line uses ``polynomial initial data'' from~\cite{Jaramillo:2020tuu,Jaramillo:2021tmt,Ansorg:2016ztf}, while orange line has Gaussian initial data with a width $M_{\rm BH}$. The inset shows the effective potential experienced by GWs, with $\epsilon = 0.1$~\eqref{eq:perturbed_potential}.}
    \label{fig:Mismatch_epsilon}
\end{figure}
The QNM spectrum of BHs is unstable under small perturbations or couplings to matter~\cite{Nollert:1996rf,Cardoso:2024mrw}. Our case study contains instabilities for the fundamental mode, and we concluded for their nonobservability. We now want to show parenthetically that high-frequency, spectrally unstable perturbations are probably stable insofar as observations go.
We stray from the galactic geometry \eqref{eq:HernquistMass}, and consider instead the toy model of Ref.~\cite{Jaramillo:2020tuu,Jaramillo:2021tmt}, where the effective potential for GW propagation gets deformed by
\begin{equation}\label{eq:perturbed_potential}
\delta V=\left(1-\frac{2M_{\rm BH}}{r}\right)\frac{\epsilon}{r^2} \sin{\left(2\pi k \frac{2M_{\rm BH}}{r}\right)}\,,
\end{equation}
where we fix $k = 10$. Despite the presence of spectral instability (in high overtones)~\cite{Jaramillo:2020tuu,Jaramillo:2021tmt}, the faithfulness decreases with $\epsilon^{2}$ as shown in Fig.~\ref{fig:Mismatch_epsilon}:~it is ``spectrally stable''. There will be no impact on observations.

\noindent {\bf \em  Discussion.}
BH spectroscopy is an indispensable tool for studying astrophysical and fundamental properties of BHs. With the increased sensitivity of future GW detectors, it will become possible to probe ringdown more accurately, and in different frequency ranges. This opens the interesting possibility of using BH ringdown to probe environments, a prospect made even more exciting with the discovery that the BH spectrum is unstable~\cite{Nollert:1996rf,Nollert:1998ys,Leung:1997was,Barausse:2014tra,Jaramillo:2021tmt,Cardoso:2019rvt,Jaramillo:2020tuu,Cheung:2022rbm,Konoplya:2022pbc,Konoplya:2022hll,Cardoso:2022whc,Cardoso:2024mrw,Yang:2024vor,Ianniccari:2024ysv}. In this work, we studied this question, using a model that resembles dark matter densities in typical galactic environments. For realistic values of environment parameters, we find that the leading-order effect is simply a (gravitational) redshift of the fundamental frequency. This may be thought of as a propagation effect, as the GW climbs out of the gravitational potential of matter. Our results are consistent with a high-frequency approximation to QNMs~\cite{Cardoso:2021wlq}, which can be used to argue that the compactness controls the QNMs of BHs in environments~\cite{Duque:2023seg}.

We studied only nonspinning BHs, possibly a conservative approach, since spin may add one more degeneracy knob on the search.

At leading order and large SNR, the error on the measurement of the BH mass yields~\cite{Berti:2005ys}
\begin{equation}
\frac{\sigma_{M_{\rm BH}}}{M_{\rm BH}}\approx \frac{2}{\mathrm{SNR}}\,,
\end{equation}
suggesting that, even if the BH mass is known \textit{a priori}, e.g.,~from inspiral-only analysis or electromagnetic counterparts, $\mathrm{SNR}\sim a_{\rm H}/M_{\rm H}$ is required to distinguish an event from vacuum. For galactic environments, this requires unreasonably loud events~\cite{Berti:2016lat,Seoane:2021kkk}. The take-home message is that BH quality factors are too small for environments to significantly impact spectroscopy.

Atomic and molecular spectroscopy in environments is well-understood. Among others, the Stark effect contributes to a distortion of spectral Balmer lines in plasmas~\cite{PhysRev.185.140,PhysRevA.6.1132,PhysRevA.11.1854}.
Environments can, in principle, also affect BH spectroscopy, but our results suggest only those with extreme density or compactness could lead to detectable effects with planned detectors.
Nevertheless, if used in conjunction with measurements of the inspiral, one may use this to our advantage to e.g.~remove redshift degeneracies or obtain information on environmental properties through pre/post merger consistency tests. As seen in Fig.~\ref{fig:Plunge_fluid}, environments do affect the very late time behavior of coalescences, possibly amplifying tail amplitudes~\cite{Albanesi:2023bgi,Carullo:2023tff,Cardoso:2024jme,DeAmicis:2024not}, a topic which deserves further scrutiny.

\noindent {\bf \em Acknowledgments.} 
We are indebted to Rodrigo Panosso Macedo for sharing the data from Ref.~\cite{Jaramillo:2021tmt}. We acknowledge support by VILLUM Foundation (grant no. VIL37766) and the DNRF Chair program (grant no. DNRF162) by the Danish National Research Foundation.
V.C.\ is a Villum Investigator and a DNRF Chair.  
V.C. acknowledges financial support provided under the European Union’s H2020 ERC Advanced Grant “Black holes: gravitational engines of discovery” grant agreement no. Gravitas–101052587. 
Views and opinions expressed are however those of the author only and do not necessarily reflect those of the European Union or the European Research Council. Neither the European Union nor the granting authority can be held responsible for them.
G.C. acknowledges funding from the European Union’s Horizon 2020 research and innovation program under the Marie Sklodowska-Curie grant agreement No. 847523 ‘INTERACTIONS’.
This project has received funding from the European Union's Horizon 2020 research and innovation programme under the Marie Sklodowska-Curie grant agreement No 101007855 and No 101131233.
\bibliography{References3}
\clearpage
\renewcommand{\thesubsection}{{E.\arabic{subsection}}}
\setcounter{section}{0}
\section*{End matter}
\subsection{Numerical setup:~a plunging particle}\label{sec:numeric_procedure}

We study the response from the BH surrounded by the environment with BH perturbation theory for non-vacuum spacetimes~\cite{Barack:2018yvs, Pound:2021qin, Cardoso:2022whc, Duque:2023seg,Cardoso:2021wlq}, where both the metric and the anisotropic fluid characterizing the halo are perturbed at linear order in the small mass ratio $q = m_{\rm p}/M_{\rm BH}$. Due to the spherical symmetry of the background, perturbations can be divided in two decoupled classes, \textit{axial} and \textit{polar}, depending on how they behave under parity transformations~\cite{Regge:1957td,Zerilli:1970wzz,Zerilli:1970se}. In this work, we are concerned with the polar sector as 
it is here that the matter and gravity perturbations are coupled.

Details on the derivations of the equations of motion governing the evolution of the perturbations can be found in~\cite{Cardoso:2022whc}. Schematically, they are given by a set of 3 wave-like equations
\begin{equation}\label{eqn:wavelikeeqn}
\hat{\mathcal{L}}\vec{\phi} = \hat{A}\vec{\phi}_{,r_{*}} + \hat{A}\vec{\phi} + \vec{S}^{\rm p}\,,
\end{equation}
where $\vec{\phi} = (S, K, \delta \rho)$, $S$ and $K$ are functions representing perturbations of the metric and $\delta \rho$ is the perturbation of the matter density profile.  $\mathcal{L}_{v} = v^{2}\partial^{2}/\partial r_*^{2} - \partial^{2}/\partial t^{2}$ is a wave operator and $\hat{\mathcal{L}}\vec{\phi} = \left( \mathcal{L}_1 \phi_1,  \mathcal{L}_1 \phi_2,  \mathcal{L}_{c_{\rm s_{r}}} \phi_3 \right)$. The coefficient matrices of the homogeneous part $\hat{A}$ and $\hat{B}$ are the same as in~\cite{Cardoso:2022whc}, with the difference in the setup being the source term $\vec{S}^{\rm p}$, representing the radial plunge of the small body. 

The latter is represented by a point-particle with world-line $x_{\rm p}^\mu(\tau)$, four-velocity $u_{\rm p}^\mu = \mathrm{d}x_{\rm p}^\mu/\mathrm{d}\tau$ and stress-energy tensor:
\begin{equation}
T_{\rm p}^{\mu \nu}=m_{\rm p} \int u_{\rm p}^\mu u_{\rm p}^\nu \frac{\delta^{(4)}\left(x^\mu-x_{\rm p}^\mu(\tau)\right)}{\sqrt{-g}} \mathrm{d} \tau\,,
\end{equation}
where $\tau$ is the proper time related to coordinate time via
\begin{equation}
\frac{\mathrm{d}t}{\mathrm{d}\tau} = \frac{E_{\rm p}}{\sqrt{A(r_{\rm p})}}\,.
\end{equation}
Here, $E_{\rm p}$ is the energy of the point particle~\cite{Cardoso:2022whc} and $A(r)$ the $g_{tt}$ component of the metric, see Eq.~\eqref{eqn:line_element}. We do not consider backreaction on the orbit due to GW emission, i.e., we assume the particle follows geodesic radial motion, determined by 
\begin{equation}\label{eqn:plunge}
\frac{\mathrm{d}r_{\rm p}}{\mathrm{d}t} = -\sqrt{A(r_{\rm p})B(r_{\rm p})}\sqrt{1 - \frac{A(r_{\rm p})}{E_{\rm p}^2}}\,,
\end{equation}
where $B(r) = 1 - 2m(r)/r$. Since we take the plunge to be along the radial $\hat{z}$--direction, such that $\theta_{\rm p} = \varphi_{\rm p} = 0$, only $m = 0$ modes are excited.

Following the same procedure as~\cite{Cardoso:2022whc}, we find $\vec{S}^{\rm p} = q \left(S_1^{\rm p}, S_2^{\rm p}, S_3^{\rm p}  \right)$ with coefficients:
\begin{equation}
\begin{aligned}
S_1^{\rm p} &=-\frac{8\sqrt{\pi}}{E_{\rm p}} \sqrt{2\ell+1}\frac{A}{r^3}\left(E_{\rm p}^2 - A\right) \sqrt{A B}\,\delta_r\,,\\
S_2^{\rm p} &= \, -\frac{4\sqrt{\pi}}{E_{\rm p}} \sqrt{2\ell+1}\frac{A}{r^2}\sqrt{AB}\, \delta_r\,,\\
S_3^{\rm p} &= -\frac{2\sqrt{\pi}}{E_{\rm p}} \sqrt{2\ell+1}\frac{\rho + 2 p_{\rm t}}{r^2}\left(2E_{\rm p}^2-A\right) \sqrt{AB}\, \delta_r\,,\\
\delta_r &\equiv \delta(r-r_{\rm p}(t))\,,        
\end{aligned}
\end{equation}
where we suppressed dependencies on $r$. Numerically, the Dirac delta representing the point particle is approximated by a smoothed Gaussian-like distribution. 
\begin{equation}
\delta(x-x_{\rm p}(t)) =\frac {\exp\left[ -(x-x_{\rm p}(t))^2/2\lambda_{\rm p}^2 \right]}{\sqrt{2\pi}\lambda_{\rm p}}\,,
\end{equation}
where $\lambda_{\rm p}$ is sufficiently small to ensure numerical convergence as $\lambda_{\rm p} \rightarrow 0$. Typically, we take $\lambda_{\rm p} \approx 4 \mathrm{d}x$, where $\mathrm{d}x$ is the grid discretization step.

The time-domain code follows earlier work~\cite{Cardoso:2022whc,Cannizzaro:2024yee} and has a uniformly spaced grid in tortoise coordinates $r_*$
\begin{equation}
\frac{\mathrm{d}r_*}{\mathrm{d}r}=\frac{1}{\sqrt{AB}}\,.
\end{equation}
To avoid boundary effects spoiling the numerical evolution, we place the inner and outer boundary sufficiently far away such that they cannot affect the system at the extraction radius. The evolution equations~\eqref{eqn:wavelikeeqn} are then integrated with a two-step Lax-Wendroff algorithm that uses second-order finite differences~\cite{Zenginoglu:2011zz}. Due to divergences at the horizon, we cut off the matter distribution and the sound speed at some radius $r_{\rm cut}$. We use $r_{\rm cut} = 2.1 M_{\rm BH}$, but we checked the qualitative results are independent of this. For initial data, we set all perturbations to $0$, and apply a window function to $\vec{S}^{\rm p}$ to start the particle smoothly, and reduce the initial junk radiation.  

From the evolved variables $\vec{\phi}$ \eqref{eqn:wavelikeeqn}, we can construct the Zerilli-Moncrief~\cite{Zerilli:1970wzz} function in the near-vacuum region as
\begin{equation}
\begin{aligned}
\mathcal{Z}_{\ell m}&=\frac{r}{n+1}\Biggl[K(r)+\frac{A(r)}{n+3(M_{\rm BH}+M_{\rm H})/r}\\
&\times \left(H_2(r)-r \frac{\partial K}{\partial r}\right)\Biggr]\,,
\end{aligned}
\end{equation}
where $n = \ell(\ell+1)/2 - 1$ and $H_2(r)$ can be found in Eq.~(47) in~\cite{Cardoso:2022whc}. This function controls the radiative degrees of freedom of the gravitational field and is related to the GWs polarizations by
\begin{equation}
\left(h_{+}-i h_{\times}\right)_{\ell m}=\frac{1}{r} \sqrt{\frac{(\ell+2) !}{(\ell-2) !}} \mathcal{Z}_{\ell m} \> {}_{{\scalebox{0.65}{$-$}}2}\mkern-2mu Y_{\ell m}(\theta, \varphi)+\mathcal{O}\left(\frac{1}{r^2}\right)\,,
\end{equation}
where ${}_{{\scalebox{0.65}{$-$}}2}\mkern-2mu Y_{\ell m}(\theta, \varphi)$ are the spin-weighted spherical harmonics with $s = 2$. For radial plunges, the cross-polarization $h_{\times}$ is zero and the GW radiation is fully determined by $h_+$.
\subsection{Waveform alignment and extrapolation}\label{sec:waveform_alignment}
Due to the slow polynomial decay of the halo mass (see Eq.~\eqref{eq:HernquistMass}), it is important to extract the signal far outside the halo, i.e., $r_{\rm ex} \gg a_{\rm H}$. In addition, to mitigate any systematic error arising from extracting at a finite radius, the signal is extrapolated to infinity. This procedure is outlined below.

Each GW signal is extracted at 4 different radii, namely $r_{\rm ex} = [20,40,80,100]a_{\rm H}$. At each radius, we locate the peak of the strain $h_{+}$ and we truncate the waveform roughly $\sim\!10M_{\rm BH}$ before the peak until the tail sets in. Subsequently, the waveforms are ``rescaled'' relative to the waveform extracted at the largest radius. In particular, labelling the latter as as $h_{1}$, we allow the (to-be-shifted) waveform $h_{2}$ to undergo a stretching $\alpha$ and a time shift $t_0$ according to
\begin{equation}\label{eq:shifted_waveform}
    \hat{h}_{2} = \int\!\mathrm{d}t\, e^{i\omega (t-t_0)}h(\alpha t) =e^{-i\hat{\omega}\alpha t_0}\tilde{h}_{2}(\hat{\omega})\,,
\end{equation}
where a tilde denotes the Fourier transform and $t = \hat{t}/\alpha$ and $\hat{\omega} = \omega\alpha$. The parameters $\alpha$ and $t_0$ are then determined by maximizing the faithfulness $\mathcal{F}(\hat{h}_{2}, h_{1})$~\eqref{eq:faithfulness}, i.e., $\mathop{\text{max}}_{\alpha, t_0} \mathcal{F}$. Since the waveforms are extracted at varying radii but correspond to the same halo configuration, the deviations in $\alpha$ and $t_0$ are minimal:~typically $\alpha \sim 1$ and $t_0 \sim 0$. The waveform is then extrapolated to infinity by fitting a Chebyshev polynomial $rh(t) = h_{\infty} + a_{1}(1/r) + a_{2}(1/r^2) + \cdots$, where $h_{\infty}$ represents the waveform at future null infinity. In our fit, we include the first two orders, yet an error estimate coming from the first and third order are included in Fig.~\ref{fig:Detectability_fluid}.

With the waveforms extrapolated to infinity, we can now compare the ringdown signal from different halo configurations. Conform to realistic GW searches, we again apply the transformation from Eq.~\eqref{eq:shifted_waveform}, taking the reference waveform $h_{1}$ to be the vacuum one. In this context, $\alpha$ acquires a clear physical meaning:~it represents the overall redshift induced by the presence of the halo. 
\end{document}